\newcommand{\gn}{\mbox{$\gamma_{\stackrel{}{5}}$}}
\newcommand{\adag}{a^{\dagger}_{p,s}}

\newcommand{\bdag}{b^{\dagger}_{-p,s}}

\newcommand{\adagbdag}{a^{\dagger}_{p,s} b^{\dagger}_{-p,s}}
\newcommand{\aps}{a^{}_{p,s}}
\newcommand{\bps}{b^{}_{-p,s}}

\newcommand{\Bdag}{B^{\dagger}_{-p,s}}
\newcommand{\Aps}{A^{}_{p,s}}

\newcommand{\psibar}{\bar{\psi}}
\newcommand{\psibarpsi}{ < \bar{\psi} \, \psi
            > }

\newcommand{\Q}{Q_{_{5}}}

\newcommand{\thetap}{\theta_{p}}
\newcommand{\costhetap}{\cos{\thetap}}
\newcommand{\sinthetap}{\sin{\thetap}}

\newcommand{\Tprime}{T'}
\newcommand{\Tprimesq}{T^{'2}}

\newcommand{\x}{\vec{x},t}

\documentstyle[12pt]{article}

\def\be{\begin{equation}}
\def\ee{\end{equation}}
\def\bea{\begin{eqnarray}}
\def\eea{\end{eqnarray}}

\begin{document}
\begin{center}
{\Large {\bf Chirality in the Early Universe }}


Ngee-Pong Chang ({ \em npccc@cunyvm.cuny.edu\ }) \\

Department of Physics \\
City College \& The Graduate School of City University of New York\\
New York, N.Y. 10031

\end{center}

\abstract{ \em
        The early big bang is an alphabet soup of quarks,
	W bosons, gluons, and other exotic particles and flavors.
        In the usual scenario, there is no place for the pion.  It
	dissociates in the alphabet soup of the
	early universe.  I will show that this scenario is naive.
        The thermal vacuum is a far more complex state, and
	the pion remains a Nambu-Goldstone particle at high $T$, and will not
        dissociate.  It propagates at the speed of light but {\em with a halo}.
}

\section{Introduction}

	In the usual folklore, chiral symmetry is restored in the early
	universe, and the pion is no longer a Nambu-Goldstone boson.
	The pion dissociates into massless quark and antiquark pairs.

	An important part of the usual folklore revolves around the
	order parameter $\psibarpsi$.  At zero temperature, $\psibarpsi$
	is non-vanishing.  The well-known theorem says that therefore
	the ground state is not chirally invariant.  A corollary of this
	theorem is that for a chiral symmetric theory,
	there must then exist a massless Nambu-Goldstone
	particle, which we know as the pion.  The fact that the physical pion
	we observe is not quite massless may be attributed to the presence of
	electroweak breaking in the Standard Model giving rise to
	a small primordial mass of the quarks.

	What happens at high $T$ ?

	Studies have shown that $\psibarpsi$ actually vanishes above a
	certain critical temperature, $T_c$.  By analogy with the
	ferromagnetic system, the usual conclusion is drawn that a
	vanishing $\psibarpsi$ indicates a chiral symmetry of the high
	temperature vacuum.
	{\em This conclusion is false, as I will show in an example. (See
	eq.(\ref{eq-new-vac}) below)  }

	However, I hasten to add, this does not mean that there was
	no phase transition taking place at $T_c$.  On the contrary, there
	is a very interesting new phase transition taking place.  It is a
	morphosis of the old zero temperature chirality.
	The original NJL vacuum undergoes an
	interesting {\em  new phase transformation} such that $\psibarpsi$
	vanishes, but the vacuum continues to break our zero temperature
	chirality.

	Above $T_c$, a new chiral symmetry takes over.
	The pion remains a Nambu-Goldstone boson,
	and actually acquires a halo while propagating through the early
	universe.

\section{High Temperature Effective Action}
	At high temperatures, lattice work as well as continuum field theory
	calculations show that the effective action indeed exhibits a manifest
	chiral symmetry. In thermal field theory, there is
	the famous Braaten-Pisarski Frenkel-Taylor-Wong (BPFTW)
	action~\cite{BP}
	that describes the
	propagation of a QCD fermion through a hot medium
	($T^{'2}  \equiv \frac{\textstyle g_r^2 }{\textstyle 3} T^2$,
	while the angular brackets denote an average over the orientation
	$\hat{n}$)
\begin{equation}
   {\cal L}_{\rm eff} = - \psibar \gamma_{\mu} \partial^{\mu}
                          \psi
                       - \frac{T^{'2}}{2\;\;} \, \psibar
                         \left<
                       \frac{\gamma_o - \vec{\gamma} \cdot \hat{n} }
                       {D_o + \hat{n} \cdot \vec{D} }
                         \right> \psi          		\label{eq-BP-action}
\end{equation}
	and we see the global chiral symmetry of the action.  But the
	{\em  nonlocality} of the action implies that the Noether charge
	for this new chirality is not the same as the usual zero temperature
	chirality.

	The fermion propagator~\cite{Weldon-Klimov} that results from
	this action shows a pseudo-Lorentz invariant particle pole of
	mass $\Tprime$ (the so-called thermal
	mass) ~\cite{Donoghue-Chang-hiT-Barton}.  But, in addition,
	there is a pair of conjugate {\em  spacelike} plasmon cuts in the
	$p_o$-plane that run just above and below the real
	axis~\cite{Chang-xc}, from $p_o = -p$ to $p_o = p$.  The cuts
	are associated with the logarithms that result from the angular
	average in eq.(\ref{eq-BP-action}).  Along the real $p_o$
	axis, the propagator function has been chosen real. As a result,
	for $t>0$, say, the propagator function takes the form
\begin{eqnarray}
	&& < T( \psi(x) \bar{\psi}(0) ) >_{_{\beta}}
	=	\;\;\; \int \frac{d^3 p}{ (2\pi)^3 } \;
		{\rm e}^{i \vec{p} \cdot \vec{x}} \;\left\{
		Z_{p} \frac{-i \vec{\gamma}
			\cdot \vec{p} + i \gamma_o \omega }{2 \omega} \;
		{\rm e}^{ - i \omega t} \right.\nonumber\\
	& &	\left. \;\;\;\;\;\;\; -  \frac{\Tprimesq}{8\;\;} \;
		\int_{-p}^{p} \,\frac{dp_o'}{p^3}  \;
		\frac{i \vec{\gamma} \cdot \vec{p} p_o'
		- i \gamma_o p^2}{p^2 - p_o^2 + \Tprimesq}   \;
		{\rm e}^{- i p_o' t}
		+ O (T'^4)
		\right\}			\label{eq-spacelike-cut}
\end{eqnarray}
	Note that the spinor structure of the massive particle pole term
	has the (unusual) feature of being manifestly chiral invariant.
	Here $Z_p$ is a wave function renormalization
	constant~\cite{Chang-bp-local}.

	In a recent study of the spacetime quantization of the
	BPFTW action~\cite{Chang-bp-local}, I have shown that
	the spacelike cuts dictate a new thermal vacuum of the
	type
\begin{equation}
	| vac' > \;=\; \prod_{p,s} \left( \costhetap
		 \;-\; i\, s \,
			\sinthetap \, \adagbdag
			\right) \; | 0 >	\label{eq-new-vac}
\end{equation}
	The $90^{o}$ phase here in the generalized NJL vacuum is
	the reason why $\psibarpsi$ vanishes for $T \geq
	T_c$.

	The quantization of a nonlocal action is of course a technical
	matter.  Suffice it here to say that the quantization has been
	formulated in terms of auxiliary fields so that the resulting action
	is local. In this context, the pseudo-Lorentz particle pole is
	described in terms of the massive canonical Dirac field, $\Psi$,
	and the spacelike cuts are associated with the auxiliary fields, which
	are functions of $\Psi$.
	This formulation allows for a systematic expansion
	of the $\psi$ field in terms of the massive canonical Dirac field,
	$\Psi$.
	Let the $t=0$ expansion for the original massless $\psi$ field read
\begin{equation}
	\psi (\vec{x}, 0) = \frac{1}{\sqrt{V}}  \sum_{p}
		\; {\rm e}^{i \vec{p} \cdot \vec{x}}
		\left\{ \left( \begin{array}{r}
			\chi_{_{p,L}} a^{}_{p,L}  \\
			\chi_{_{p,R}} b^{\dagger}_{-p,R}
			\end{array}\right)
		 \;+\;  \left( \begin{array}{r}
			\chi_{_{p,R}} a^{}_{p,R} \\
			\;-\;
			\chi_{_{p,L}} b^{\dagger}_{-p,L}
			\end{array} \right) \right\}
\end{equation}
	with a corresponding canonical expansion for the massive $\Psi$,
	then we find
\begin{eqnarray}
	\aps	&=& \Aps \;-\; i \;s\; \frac{\Tprime}{2 p}
		    \Bdag + O(\Tprime {}^2 )	\label{eq-aps-Aps-1} \\
	b^{}_{p,s} &=& B^{}_{p,s} \;+\; i \;s\; \frac{\Tprime}{2 p}
		    A^{\dagger}_{-p,s}
			+ O(\Tprime {}^2)	\label{eq-bps-Bps-1}
\end{eqnarray}
	The $O(\Tprime)$ terms in the Bogoliubov transformation
	imply the new thermal vacuum of eq.(\ref{eq-new-vac}).

	The chiral charge at high $T$ is given by
\begin{equation}
        Q_{5}^{\beta} =  - \frac{1}{2} \; \sum_{p,s}\; s \;
                      \left(
                      A^{\dagger}_{p,s} A^{}_{p,s} + B^{\dagger}_{-p,s}
                      B^{}_{p,s}
                                  \right)
\end{equation}
	so that it annihilates the new thermal vacuum,
        in direct contrast with the $T=0$ Noether charge
\begin{equation}
        Q_{_{5}}
        = - \frac{1}{2}
             \sum_{p,s}\, s \; \left( a^{\dagger}_{p,s} a^{}_{p,s} +
             b^{\dagger}_{-p,s} b^{}_{-p,s}
			\right)			\label{Q5}
\end{equation}
	which clearly fails to annihilate the vacuum at high $T$.

\section{$\psibarpsi$ is an Incomplete Order Parameter}

	The traditional order parameter $\psibarpsi$
	cannot by itself give a full description of the nature
	of chiral symmetry breaking.  The operator, $\bar{\psi} \psi$,
	belongs to a non-Abelian chirality algebra,
	$SU(2N_f)_{p} \otimes SU(2N_f)_{p}$.
	This algebra is not to be confused with the $U(1)_{_{V}}
	\otimes SU(N_f)_{_{L}}
	\otimes SU(N_f)_{_{R}}$ algebra that is a symmetry of the
	fundamental massless Lagrangian. This symmetry is broken
        spontaneously by the vacuum to the surviving vector symmetry
	$U(1)_{_{V}} \otimes SU(N_f)_{_{V}}$.

	The $SU(2N_f)_{p} \otimes SU(2N_f)_{p}$ chirality algebra we
	refer to here is instead
	a spectrum generating algebra, so that in general elements
	of the algebra do not commute with the Hamiltonian.
	They are nevertheless useful in classifying the properties
	of the dynamical vacuum that result.

	The original chiral broken NJL ground state
	may be written as an $X_2$ rotation of the usual Fock vacuum
\begin{equation}
	|vac> = \prod_{p} {\rm e}^{i X_{2p} \thetap}\; |0>
\end{equation}
	where $X_{2p}$ is an element of the algebra, given in terms of the
	massless quark operators by
\begin{equation}
	X_{2p}	=	\,i \,\sum_s \; \; \frac{s}{2}  \left( \adag \bdag
			- \bps \aps \right)
\end{equation}
	while
	the new thermal vacuum eq.(\ref{eq-new-vac}) is generated by a
	different element, $Y_{1p}$,
\begin{equation}
	Y_{1p}	=	\;-\;\; \sum_s \;\, \frac{s}{2} \,\,\left( \adag
			\bdag  + \bps \aps \right)
\end{equation}
	It is interesting to note that the usual order parameter is
	related to this element through the volume integral
$
	\frac{1}{2} \int d^3 x  \; \bar{\psi}(\vec{x}) \; \psi (\vec{x})
		\;=\;
		- \sum_{p} Y_{_{1p}}.
$
	The remaining elements of the algebra exhaust the bilinears that may
	be formed from the massless quark operators
\begin{eqnarray}
     	X_{3p}	&=&	- \sum_s \; \frac{s}{2}  \;\left( \adag \aps
			+ \bdag \bps \right) \\
	X_{1p}	&=&	\;\; \,\sum_s \;\, \frac{1}{2} \,\,\left( \adag \bdag
			+ \bps \aps \right) \\
	Y_{1p}	&=&	\;\; \sum_s \;\, \frac{s}{2} \,\,\left( \adag
			\bdag  + \bps \aps \right) \\
     	Y_{3p}	&=&	\;\; \sum_s \; \frac{1}{2}  \;\left( \adag
			\aps  - \bps \bdag \right)
\end{eqnarray}
	The $X$ operators generate the $SU(2)$ algebra, with $X_{3p}$
	easily recognizable as the Fourier component of the usual chirality
	charge,
$
	Q_{5} \;=\;  \sum_{p} \; X_{3p}.
$
	While $X_{3}, Y_{1}$ are related to local operators in coordinate
	space, the other elements are related to necessarily nonlocal
	operators.
	Our results here suggest the study of a new class of these
	nonlocal order parameters that involve a time integration
	that projects away the usual timelike spectrum of the operator
	$\psi$, and probes directly the properties of the spacelike
	cut.

\section{Pion halo in the Sky}

	The pion we know at zero temperature is not massless,
	but has a mass of $135 \;MeV$.  This is because of
	electroweak breakdown, giving rise to a primordial quark mass at
	the tree level.
	At very high $T$, when electroweak symmetry is restored, we
	have the interesting new possibility that the pion will fully
	manifest its Nambu-Goldstone nature and remain physically
	massless.~\cite{Chang-QCD}

	The pion is described by an interpolating field operator,
$
	 \pi^{a}  \;\sim\;   i \bar{\psi} \gn T^{a} \; \psi,
$
	which does not know about temperature. It is the vacuum that
	depends on $T$.
	The state vector for a zero momentum pion at high $T$ may
	be obtained from the thermal vacuum by the action
\begin{equation}
	\Q^{a} | vac' > \;\propto \; | \pi^{a} ( \vec{p} = 0 ) \rangle
\end{equation}
	This pion now has the property that even though it is
	massless, it can acquire a {\em  screening mass} proportional
	to $T$.
	This is the pion mass that has
	been measured on the lattice at high $T$.

	As a result, the pion propagates in the early universe with a
	halo.  The retarded function for the pion shows that the
	signal propagates along the light cone, with an additional
	exponentially damped component coming from the past history
	of the source.
\begin{eqnarray}
        D_{\rm ret} (\x) &=& \theta (-t) \left\{ \delta(t^2 - r^2)
			\right. \nonumber \\
                    	& & + \frac{\Tprime}{r} \theta(t^2 - r^2)
                    \left[ {\rm e}^{-\Tprime | t-r| }
                    	 +  \left. {\rm e}^{-\Tprime | t+r| }
				\right] \right\}
\end{eqnarray}
        The screening mass leads to an accompanying modulator
        signal that `hugs' the light cone, with a screening length
        $\propto 1/T$.

	What are the cosmological consequences of a pion in the
	alphabet soup of the early universe?

	In the usual scenario, the pion after chiral restoration
	will have acquired mass $\propto T$, and will quickly dissociate
	into constituent quark-antiquark pair.  According to our
	new understanding, however, the Nambu-Goldstone theorem forces
	the pion to remain a strictly massless bound state at high $T$,
	and so the pion will contribute to the partition function
	of the early universe.

	Fortunately, the pion does not contribute so many degrees of freedom
	as to upset the usual picture of the cooling of the universe.  But
	I leave it to experts to help figure out the subtle changes there
	must surely be in the phase transitions of the early universe.

	{\em In the beginning there was light, and quarks, and gluons, to which
	we must now add the pions with halo. }

\end{document}